\documentclass[12pt]{iopart}
\usepackage{iopams}  

\makeatletter
\def\cite{\@ifnextchar[{\@tempswatrue\@citex}{\@tempswafalse\@citex[]}}

\def\@citex[#1]#2{%
  \if@filesw\immediate\write\@auxout{\string\citation{#2}}\fi%
  \leavevmode\unskip\ \@cite{\@collapse{#2}}{#1}}%

\def\@cite#1#2{[{#1\if@tempswa , #2\fi}]} %

\def\@collapse#1{%
{%
\let\@temp\relax
\@tempcntb\@MM
\def\@citea{}%
\@for \@citeb:=#1\do{%
\@ifundefined{b@\@citeb}%
{\@temp\@citea{\bf ?}%
\@tempcntb\@MM\let\@temp\relax
\@warning{Citation `\@citeb ' on page \thepage\space undefined}%
}%
{\@tempcnta\@tempcntb \advance\@tempcnta\@ne
\edef\MyTemp{\csname b@\@citeb\endcsname}%
\def\@tempa{\@temptokena=\bgroup}%
\afterassignment\@tempa %
\@tempcntb=0\MyTemp\relax}%
\ifnum\@tempcntb=0\relax%
\@tempcntb=\@MM
\@citea\MyTemp
\let\@temp = \relax
\else %
\edef\@tempd{\number\@tempcntb}%
\ifnum\@tempcnta=\@tempcntb %
\ifx\@temp\relax %
\edef\@temp{\@citea\@tempd}%
\else
\edef\@temp{\hbox{--}\@tempd}%
\fi
\else %
\@temp\@citea\@tempd
\let\@temp\relax
\fi
\fi
}%
\def\@citea{,$\,$}%
}%
\@temp %
}%
}%
\makeatother
\begin{document}

\title{Critical behaviour of the Rouse model for gelling polymers}

\author{Peter M\"uller}
\ead{Peter.Mueller@physik.uni-goettingen.de} 

\address{Institut f\"ur Theoretische Physik,
  Georg-August-Universit\"at, D--37077 G\"ottingen, Germany}

\submitto{\JPA \emph{(in press)}}

\begin{abstract}
It is shown that the traditionally accepted ``Rouse values'' for the critical
exponents at the gelation transition do not arise from the 
Rouse model for gelling polymers. The true critical
behaviour of the Rouse model for gelling polymers is obtained from
spectral properties of the connectivity 
matrix of the fractal clusters that are formed by the molecules. The
required spectral properties are related to the return probability of a
``blind ant''-random walk on the critical percolating cluster. The
resulting scaling relations express the critical exponents of the
shear-stress-relaxation function, and hence those of the shear
viscosity and of the first normal stress coefficient, in terms of
the spectral dimension $d_{s}$ of the critical percolating cluster and the
exponents $\sigma$ and $\tau$ of the cluster-size distribution.
\end{abstract}



%
\section{Introduction and result}
%

During a chemical gelation reaction the crosslinking of
\mbox{(macro-)} molecules  
leads to the formation of permanent clusters of very different shapes
and sizes. In the absence of a macroscopically large cluster such a
system acts like a liquid (sol). Otherwise one deals with an amorphous
solid state (gel). Hence, the sol-gel or gelation transition appears as a
percolation transition \cite{StCo82,DaLa90,MaAd91,WiMo97}, which is
driven by crosslink concentration.

Gelling liquids exhibit striking mechanical properties which have been
continuously studied over the years by experiments
\cite{AdDe81,DuDe87,MaAd88,AdMa90,DeBo93,CoGi93,VlCh98,ToFa01},  
theories \cite{Sta76,Gen78,Cat84,MaAd89,RuZu90,BrGo97,ZiGr98,%
  BrLo99,BrLo01a,BrAs01,BrLo01b,BrMu02}  
and simulations \cite{GaAr00,VePl01}.  
For example, when subjected to a homogeneous shear flow, 
distinct relaxation patterns are observed, which are due to
the participation of many different excitation modes. More precisely,
experiments suggest the scaling form \cite{DaLa90,WiMo97,%
  DuDe87,MaAd88,AdMa90,ToFa01,MaAd89}
\begin{equation}
  \label{gscale}
  G(t) \sim t^{-\Delta} g(t/t^{*}) \hspace{1.5cm}\mathrm{with}\hspace{1.5cm}
  t^{*}(\varepsilon) \sim \varepsilon^{-z}  
\end{equation}
for the (shear-) stress-relaxation function in the sol phase for
asymptotically long times 
$t$ and crosslink concentrations $c$ close 
to the critical point $c_{\mathrm{crit}}$, \emph{i.e.}\
for $\varepsilon := (c_{\mathrm{crit}} -c)/c_{\mathrm{crit}} \ll 1$.
The typical relaxation time $t^{*}$ diverges with a critical
exponent $z>0$ for $\varepsilon\downarrow 0$. The scaling
function $g$ is of order unity for small 
arguments so that one
finds the algebraic decay $G(t)\sim t^{-\Delta}$ with a critical
exponent $\Delta>0$ for $t\to\infty$ at
the critical point $\varepsilon =0$. For large arguments $g$ 
decreases faster than any inverse power. Sometimes a stretched 
exponential has been proposed \cite{AdMa90,MaAd89} for $g$, implying the
long-time asymptotics $ - \ln G(t) \sim (t/t^{*})^{\gamma}$ with some
$0 < \gamma <1$ in the sol phase.  

Within linear viscoelasticity one finds $\eta =
\int_{0}^{\infty}\!\rmd t\, G(t)$ for the zero-frequency viscosity and 
$\Psi_{1} = 2 \int_{0}^{\infty}\!\rmd t\, tG(t)$ for the first normal stress
coefficient \cite{WiMo97,Fer80}. The
scaling form (\ref{gscale}) of the stress-relaxation function 
determines their critical divergencies $\eta\sim \varepsilon^{-k}$ and
$\Psi_{1} \sim \varepsilon^{-\ell}$ as $\varepsilon\downarrow 0$ by
the scaling relations \cite{WiMo97,BrMu02} 
\begin{equation}
  \label{scaling}
  k = z(1-\Delta) \hspace{2cm}\mathrm{and}\hspace{2cm}
  \ell = z(2-\Delta) = k+z\,.
\end{equation}
Thus, it suffices to know any two of the four critical exponents $\Delta,
z, k$ and $\ell$.

The Rouse model is one of the most basic models for polymer dynamics. 
By definition \cite{Rou53,DoEd86,BiCu87}, it is a bead-spring model
for linear polymer chains with no other interactions among monomers
(\emph{i.e.}\ beads) than connectivity along the chain, as accounted
for by harmonic springs. Moreover, each bead behaves as a Brownian
particle in that it
experiences a friction force and a thermal noise force. Despite its
simplicity the Rouse model has provided a much valued description of linear
polymer chains over the last fifty years. Its generalization from
linear chains to arbitrarily connected clusters of monomers is
straightforward and will be recalled in Section~\ref{model}. 
Static and dynamic properties of such Rouse clusters have been
discussed, for example, in \cite{Cat84,Vil88,SoSc93,SoBl95,BlJu02}, and
recent applications to gelling polymer melts can be found in 
\cite{BrGo97,BrLo99,BrLo01a,BrAs01,BrLo01b,BrMu02}.
More than twenty years earlier, de~Gennes was the first who aimed at a
description of the gelation transition in terms of the Rouse
model. Guided by scaling arguments, he claimed the relation 
$k=2\nu -\beta$ for the critical exponent of the 
viscosity \cite{Gen78}. Here, the exponent $\nu$ governs the
divergence of the correlation length and $\beta$ the growth of the
gel fraction. Using this result, it was also argued \cite{DuDe87,%
  MaAd88,CoGi93,MaAd89,RuCo89} that $\Delta = d\nu/(d\nu +k)$, where
$d$ is the space dimension.
Inserting the numerical values for
three-dimensional bond percolation \cite{StAh94,BuHa96}, as is
generally accepted, gives 
$k\approx 1.35$ and $\Delta \approx 0.66$. These results have become
known as the ``Rouse values'' for the critical exponents and are
commonly quoted in the literature, see e.g.\ the reviews \cite{StCo82,%
  DaLa90,MaAd91,WiMo97}. 

In contrast, the exact solution of the Rouse model for gelling
polymers in \cite{BrLo99,BrLo01a} leads to 
\begin{equation}
  \label{ka}
  k= (1-\tau +2/d_{s})/\sigma
\end{equation}
and thus disproves the earlier scaling arguments---and hence the Rouse 
value---for $k$ mentioned above. Here  
$d_{s}$ denotes the spectral dimension \cite{StAh94,BuHa96,AlOr82}  
of the incipient percolating cluster, and the exponents $\sigma$ and
$\tau$ are related to the cluster-size distribution, as defined in
(\ref{tauscaling}) below. The derivation of (\ref{ka}) in
\cite{BrLo99,BrLo01a} exploits a correspondence between
Rouse clusters and an electrical resistor network, which rests on
the linearity of Hooke's law and Ohm's law. This correspondence is exact.
It differs from de Gennes' superconductor/conductor-network analogy
\cite{Gen79}, as can be seen
most clearly by comparing the resulting numerical values for $k$
in two space dimensions \cite{BrLo99}. Unfortunately, the approach of
\cite{BrLo99,BrLo01a}, which led to (\ref{ka}), does not seem to be suitable
for a computation of the other exponents $\Delta$, $z$ or $\ell$---except for
the case where the statistical ensemble of the clusters corresponds to
mean-field percolation \cite{BrAs01,BrLo01b,BrMu02}. Hence, the
authors of \cite{BrMu02} had to use
the numerically estimated value $\Delta \approx 0.83$, which was obtained in
\cite{BrAs01} for the Rouse model for gelling polymers with clusters
distributed according to three-dimensional percolation.    

The present Paper fills this gap. We present a simple analytical
calculation which determines both $\Delta$ and $z$---and hence also
$k$ and $\ell$ by (\ref{scaling})---for Rouse clusters
distributed according to $d$-dimensional percolation.
This will be achieved by a well-known relation between the
dynamics of Rouse clusters and a ``blind ant''-random 
walk on percolation clusters \cite{Cat84,StAh94,BuHa96}. In
particular, we will express the stress-relaxation 
function $G$ in terms of the return probability of such a random walk.
In this way we will deduce the scaling relations   
\begin{equation}
  \label{deltaz}
  \Delta = \frac{d_{s}}{2}(\tau -1)
  \hspace{2cm}\mathrm{and}\hspace{2cm}
  z= \frac{2}{d_{s}\sigma}
\end{equation}
for the microscopic Rouse model for gelling polymers.
Of course, (\ref{deltaz}) and (\ref{scaling}) are consistent with
existing exact results, that is, equation (\ref{ka}) for $k$
\cite{BrLo99,BrLo01a} and the mean-field values for $\Delta$, $z$
and $\ell$ \cite{BrAs01,BrLo01b,BrMu02}. Table~\ref{expo} summarizes
the exact values of 
the critical exponents pertaining to the Rouse model for gelling
polymers with clusters distributed according to two- and
three-dimensional lattice-bond percolation, respectively mean-field
percolation.  

\begin{table}
  \caption{\label{expo}Critical exponents for a gelling Rouse melt
    based on (\ref{deltaz}) and (\ref{scaling}).} 
  \begin{indented}
  \item[]\begin{tabular}{@{}lllll}
      \br \lineup
      \phantom{mmm} & $\Delta = d_{s}(\tau -1)$/2 & 
      $z= 2/(d_{s}\sigma)$ & 
      $  k=z(1-\Delta) $ & $ \ell = z(2-\Delta) $ \\ \mr
      2D & 0.70 & 3.8 & 1.2 & 5.0 \\
      3D & 0.79 & 3.3 & 0.71 & 4.1 \\ 
      \textsc{mf} & 1 & 3 & 0 & 3\\
      \br
    \end{tabular}
  \end{indented}
\end{table}

The result for $\Delta$ in (\ref{deltaz}) differs 
from the relation $\Delta=d\nu/(d\nu +k)$ that has
grown out of the scaling arguments mentioned above. Hence, it is not
only the ``Rouse value'' for $k$ but also the one for $\Delta$ that
has been ascribed incorrectly to the Rouse model for gelling polymers.

The rest of the Paper is organized as follows. After recalling the
Rouse model for gelling polymers in Section~\ref{model}, we will derive
the scaling relations (\ref{deltaz}) within this model in
Sections~\ref{stress} to~\ref{critical}. Section~\ref{outlook}
contains a discussion and an outlook.

%
\section{Rouse model for gelling polymers} \label{model}
%

In order to set up the notation we recall the standard definition 
\cite{Cat84,DoEd86,BiCu87,SoSc93,SoBl95,BlJu02} 
of the Rouse model for clusters of monomers with an arbitrary
topology. Previous applications to gelling polymer melts can be found in 
\cite{BrGo97,BrLo99,BrLo01a,BrAs01,BrLo01b,BrMu02}. 
We consider $N$ 
point-like monomers with positions ${\bi{R}}_i(t)$, $i=1,\ldots,N$, in
$d$-dimensional space, $d\ge 2$. Their motion is constrained by
$M$ randomly chosen, permanent, harmonic crosslinks which connect the
pairs of monomers $(i_e,j_{e})$, $e=1,\ldots,M$, and give rise to the
potential energy
\begin{equation} \label{poten}
  U := \frac{d}{2a^2}\:\sum_{e=1}^M 
  \bigl( {\bi{R}}_{i_{e}}-{\bi{R}}_{j_{e}} \bigr)^2
  =: \frac{d}{2a^2}\:\sum_{i,j}^N \Gamma_{ij}\,
  {\bi{R}}_{i}\cdot{\bi{R}}_{j} \,.
\end{equation}
Here, we have chosen units such that the inverse temperature
$(k_{\mathrm{B}}T)^{-1}$ is equal
to one, and the fixed length $a>0$ serves as an inverse crosslink strength.
The second equality in (\ref{poten})
introduces the random $N\times N$-connectivity matrix $\Gamma$, which
encodes all properties of a given crosslink realisation.
The simple relaxational dynamics 
\begin{equation} \label{eqmotion}
 0 = - \zeta \left[ \frac{\rmd}{\rmd t} \bi{R}_i(t)
- \bi{v}_{\mathrm{ext}}\bigl(\bi{R}_i(t),t\bigr)\right] 
  - \frac{\partial U}{\partial\bi{R}_i}(t) 
  + \boldsymbol{\xi}_i(t)
\end{equation}
governs the motion of the monomers in the externally applied shear flow
$ v^{\alpha}_{\mathrm{ext}}(\bi{r},t) := \kappa(t)\,\delta_{\alpha,
  1}\, r_{2} $, which is oriented along the $1$-direction and involves
the spatially homogeneous, 
time-dependent shear rate $\kappa(t)$. Here, Greek indices label
Cartesian components, $\bi{r} =(r_{1},\ldots,r_{d})$ and
$\delta_{\alpha,\beta}$ denotes the Kronecker symbol.  A friction force with
friction constant $\zeta$ 
applies to monomer $i$, if its velocity deviates from
that of the surrounding shear flow. Moreover, the crosslinks exert the Hookean
forces $-\partial U/\partial{\bi{R}}_i$, and the
last contribution in (\ref{eqmotion}) is due to a randomly
fluctuating thermal-noise force $\boldsymbol{\xi}_{i}(t)$, obeying Gaussian
statistics with zero mean and covariance 
$\overline{\xi_{i}^{\alpha}(t)\xi_{j}^{\beta}(t')} = 2\zeta
\delta_{\alpha,\beta}\delta_{i,j}\delta(t-t')$. Here, $\delta(t)$
denotes the Dirac delta function. Given the homogeneous
shear flow $\bi{v}_{\mathrm{ext}}$ and any fixed configuration
$\mathcal{C} := \{(i_{e},j_{e})\}_{e =1}^{M}$ of permanent 
crosslinks, the equation of motion (\ref{eqmotion}) 
is linear and can be solved exactly for every realisation of the
thermal noise \cite{BrLo01a}. 

It remains to specify
the statistical ensemble which determines the probabilistic occurrence
of different crosslink configurations $\mathcal{C}$. Here we consider any  
bond-percolation ensemble which is amenable to a scaling description,
does not distinguish any site (\emph{i.e.}\ monomer), 
yields a fractal percolating cluster at criticality and where the
maximum number of bonds (\emph{i.e.}\ crosslinks) emanating from a given
site is uniformly bounded.
Besides certain versions of continuum percolation \cite{HaZw77}, this
also includes 
bond percolation on the simple cubic lattice in $d$ dimensions and on
the Cayley tree \cite{StAh94,BuHa96}. The statistical average over
all crosslink configurations will be denoted by angular brackets
$\langle\,{\bullet}\,\rangle$ and, when 
using this notation, we implicitly assume that the macroscopic limit
$N\to\infty$, $M\to\infty$, $M/N\to c$ is carried out, too.

%
\section{Stress relaxation in the Rouse model for gelling  polymers}
\label{stress}
%

The crosslinks exert shear stress on the molecular clusters due to the
externally applied shear flow. Following Kirkwood \cite{DoEd86,BiCu87}, 
the Cartesian components of the stress tensor are given in
terms of a force-position correlation 
\begin{equation}
  \label{stressdef}
  \sigma_{\alpha\beta}(t) = \lim_{t_{0}\to -\infty}
  \frac{\rho_{0}}{N}\,\sum_{i=1}^{N} \overline{\frac{\partial U}{\partial
    R_{i}^{\alpha}}(t) \; R_{i}^{\beta}(t)}\,.
\end{equation}
Here, $\rho_{0}$ denotes the density of monomers, and
one has to insert the explicitly known \cite{BrLo01a} solutions
${\bi{R}}_{i}(t)$ of 
the Rouse equation (\ref{eqmotion}) with some initial data
${\bi{R}}_{i}(t_{0})$. Finally, the limit $t_{0}\to - \infty$ ensures
that  all transient effects stemming from the initial data will
disappear. This yields the exact result 
\begin{equation}
  \label{stresstensor}
  \langle\boldsymbol{\sigma}(t)\rangle 
  = G(0) \boldsymbol{1}
   + \int_{-\infty}^{t}\!\rmd t' \, G(t-t')\, \kappa(t')
  \left(\begin{array}{ccccc}
      2\int_{t'}^{t}\!\rmd s\,\kappa(s) & 1 & 0      & \cdots \\
      1         & 0      & 0      & \cdots \\ 
      0              & 0      & 0      & \cdots \\[-.8ex]
      \vdots         & \vdots & \vdots & \ddots 
  \end{array}\right)
\end{equation}
for the crosslink-averaged, macroscopic stress tensor, where
$\boldsymbol{1}$ is the $3\times 3$-unit matrix and 
\begin{equation}
  \label{stressrelax}
  G(t) \equiv G(t,\varepsilon):=
  \left\langle \frac{\rho_{0} }{N}\; {\mathrm{Tr}} \left[ (1-E_0) \,
      \exp\left(- \frac{6t}{\zeta a^2} \Gamma\right)\right]\right\rangle
\end{equation}
represents the stress-relaxation function \cite{BrLo01a,BrMu02}. 
Here, ${\mathrm{Tr}}$ is the trace
and $E_0$ denotes the eigenprojector on the
space of zero eigenvalues of $\Gamma$, which correspond to translations
of whole clusters \cite{BrGo97,BrLo99,BrLo01a}. 
Within the simple
Rouse model the zero eigenvalues do not contribute to stress relaxation
because there is no force acting between different clusters. The only
contribution to stress relaxation is due to deformations of the
clusters.

%
\section{Cluster decomposition}
%

A crosslink configuration $\mathcal{C}$ can be decomposed into 
$K \equiv K(\mathcal{C})$ disjoint clusters, and we denote the number of
monomers in its $k$th cluster $\mathcal{N}_{k} \equiv
\mathcal{N}_{k}(\mathcal{C})$ by $N_{k} \equiv N_{k}(\mathcal{C})$. The
connectivity matrix  $\Gamma(\mathcal{C})$ of the 
whole configuration is thus a block-diagonal sum of connectivity
matrices $\Gamma(\mathcal{N}_{k})$ of different clusters.
Hence, when evaluating the trace in (\ref{stressrelax}), one gets an 
additive contribution from every cluster. Upon sorting these
contributions with respect to the number of monomers in the clusters
and introducing the mean number density $\tau_{n} := \bigl\langle  N^{-1}
\sum_{k=1}^{K} \delta_{N_{k},n}\bigr\rangle$ of clusters with $n$
monomers---also known as cluster-size distribution---, we arrive at
the cluster-decomposed form  
\begin{equation}
  \label{gdecomp}
  G(t,\varepsilon) =  \sum_{n=1}^{\infty} n \tau_{n}(\varepsilon)\,
  G_{n}(t,\varepsilon) 
\end{equation}
of the stress-relaxation function. 
Here, the mean stress-relaxation function of $n$-clusters
\begin{equation}
  \label{gn}
  G_{n}(t,\varepsilon) := \left\langle \frac{\rho_{0} }{n}\;
    {\mathrm{Tr}} \left[   (1-E_0) \, 
\exp\left(- \frac{6t}{\zeta a^2} \Gamma\right)\right]\right\rangle_{n}
\end{equation}
is defined in terms of the (conditional) average $\langle A\rangle_{n}
:= \tau_{n}^{-1} \bigl\langle N^{-1} \sum_{k=1}^{K} \delta_{N_{k},n}
A(\mathcal{N}_{k}) \bigr\rangle$ for an observable $A$ over all
clusters with $n$ monomers. The 
cluster-size distribution $\tau_{n}$ gives rise to the critical
exponents $\sigma$ and $\tau$ through its scaling form 
\begin{equation}
  \label{tauscaling}
  \tau_{n}(\varepsilon) \sim n^{-\tau} f_{1}\bigl(n/n^{*}(\varepsilon)\bigr)
\end{equation}
for $n \gg 1$ and crosslink concentrations
close to the critical point \cite{StAh94,BuHa96}. It involves the
typical cluster mass 
$n^{*}(\varepsilon) \sim \varepsilon^{-1/\sigma}$, which diverges for
$\varepsilon\downarrow 0$. The scaling function $f_{1}$ decreases faster
than any inverse power for large arguments, whereas for small
arguments it approaches a non-zero, finite constant. 
It will turn out later that in order to calculate the critical
exponents $\Delta$ and $z$ of $G$, it suffices to know $G_{n}$
at the gelation transition. For this reason we will only be
concerned with critical percolation clusters in the next paragraph.

%
\section{Blind ant on critical percolation clusters}
%

Consider a random walker---coined ``blind ant'' by de~Gennes
\cite{Gen76}---that moves along a bond from one site to another in the
same cluster at discrete time steps
\cite{StAh94,BuHa96,AlOr82,HaBe02}. 
More precisely, let $b$ denote the maximum 
number of possible bonds which are allowed to emanate from a site in
the percolation 
model under consideration. If the ant happens to visit site $i$ at
time $s$, which is connected with $b_{i} \le b$ bonds to other
sites, then it will move with equal probability $1/b$ along any one
of the $b_{i}$ bonds within the next time step and stay at site $i$
with probability $1 - b_{i}/b$. By the definition (\ref{poten}) of the
connectivity matrix $\Gamma$ of the cluster, one has $\Gamma_{ii} =
b_{i}$ for its diagonal matrix elements, $\Gamma_{ij} = -1$ if two
different sites $i \neq j$ are connected by a bond and zero
otherwise. Hence, the associated master equation for the
ant's sojourn probability $p_{i}(s)$ for site $i$ at time $s$ reads 
\begin{equation}
  p_{i}(s+1) = (1- \Gamma_{ii}/b) p_{i}(s) + \sum_{j \neq i}
  (-\Gamma_{ij}/b) p_{j}(s)\,,
\end{equation}
which is equivalent to
\begin{equation}
  \label{master}
  p_{i}(s+1) -p_{i}(s) = - b^{-1} \sum_{j} \Gamma_{ij} p_{j}(s)\,.
\end{equation}
Here the
summation extends over all sites in the cluster. For long times $s
\gg 1$, it is legitimate to replace the difference (quotient) on the
left-hand side of (\ref{master}) by a derivative. This yields the
solution $p_{i}(s) = \bigl[\rme^{-s\Gamma/b}\bigr]_{ii_{0}}$, which
corresponds to the initial condition $p_{i}(0) = \delta_{i,i_{0}}$.

Next we consider $P^{(n)}(s) := \left.\langle p_{i_{0}}(s)
\rangle_{n}\right|_{\varepsilon=0}$, the mean return probability to the
starting point after time $s$, where the average is taken over all
critical percolation clusters with $n$ sites. Likewise,
$P^{(\infty)}(s)$ stands for the mean return probability on the
incipient percolating cluster. Clearly, these definitions are independent
of the starting point $i_{0}$, because on average there is no
distinguished site by assumption. Thus we can also write 
\begin{equation}
  \label{nreturn}
  P^{(n)}(s) = \left. \left\langle \frac{1}{n} \; {\mathrm{Tr}}\;
    \rme^{-s\Gamma/b} \right\rangle_{n} \right|_{\varepsilon =0}
\end{equation}
for finite $n$. 
The decay of $P^{(\infty)}(s)$ for long times $s \gg 1$ is 
determined by the inverse number of distinct sites which the ant has visited
up to time $s$ and, thus, by the spectral dimension $d_{s}$ of the incipient 
percolating cluster according to \cite{BuHa96,AlOr82,HaBe02}
\begin{equation}
  \label{infreturnas}
  P^{(\infty)}(s) \sim s^{-d_{s}/2} \,.
\end{equation}

The fractal nature of percolation clusters at the critical
point implies that clusters with $n \gg 1$ sites typically look like
regions of $n$ connected sites that are cut out of the infinite
cluster. Therefore we conclude $P^{(n)}(s) \sim s^{-d_{s}/2}$ for
times $ 1 \ll s \lesssim s_{n}$, where \cite{BuHa96,AlOr82,HaBe02}
\begin{equation}
  \label{sn}
  s_{n} \sim n^{2/d_{s}}
\end{equation}
is the crossover time at which the ant first recognizes that it does
not move on 
the infinite cluster, but on a finite one with $n$ sites,
\emph{i.e.}\ the time it needs to travel a distance that is comparable to the
cluster's extension. For times $s \gg s_{n}$, on the other hand, every site
in the cluster is equally likely to be visited by the ant, hence
$P^{(n)}(s) \sim 1/n$ in this case. Taken together, we can write
\begin{equation}
  \label{nreturnas}
  P^{(n)}(s) \sim s^{-d_{s}/2} f_{2}(s/s_{n}) +1/n
\end{equation}
for $s \gg 1$. The cut-off function $f_{2}(x)$ is of order one for $x
\lesssim 1$ and decreases rapidly to zero for $x \to\infty$. Upon
comparing (\ref{nreturn}) and (\ref{nreturnas}), we infer the relation 
\begin{equation}
  \label{gnas}
  \left. \left\langle \frac{1}{n}\; {\mathrm{Tr}}\,
    \bigl[(1-E_{0}) \,\rme^{-s\Gamma/b} \bigr] \right\rangle_{n}\;
  \right|_{\varepsilon =0} \sim s^{-d_{s}/2} f_{2}(s/s_{n})
\end{equation}
for $s \gg 1$. To derive (\ref{gnas}) we have also used the explicit
form $(E_{0})_{ij} =n^{-1}$, $1 \le i,j \le n$, of the matrix elements of the
eigenprojector corresponding to the non-degenerate zero eigenvalue of
$\Gamma$ for the case of a single cluster with $n$ sites.

%
\section{Critical behaviour of stress relaxation} \label{critical}
%

The long-time decay of the stress relaxation function $G$ is
determined by the biggest clusters. Close to the gelation transition
we may thus use the scaling form (\ref{tauscaling}) of the
cluster-size distribution and approximate the (Riemann) sum on the
right-hand side of (\ref{gdecomp}) by the corresponding integral
\begin{equation}
  \label{gint}
  G(t,\varepsilon) \sim  (n^{*}(\varepsilon))^{2-\tau} \int_{0}^{\infty}
  \!\rmd x\; x^{1-\tau} f_{1}(x)\,
  G_{xn^{*}(\varepsilon)}(t,\varepsilon) \,.
\end{equation}
If (\ref{gint}) is to be consistent with the scaling form
(\ref{gscale}), we are forced to require the functional dependence
\begin{equation}
  \label{gneps}
  G_{n}(t,\varepsilon) \sim t^{-a} f_{3}\bigl( n/n^{*}(\varepsilon),
  t/t^{*}(\varepsilon)\bigr) \,.
\end{equation}
The yet unknown exponent $a$ and the time scale $t^{*}(\varepsilon) \sim
\varepsilon^{-z}$  can be determined from the limiting case
$\varepsilon =0$,
\begin{equation}
  \label{gn0}
  G_{n}(t,0) \sim t^{-d_{s}/2} f_{2}(t/t_{n})
  \qquad\mathrm{with}\qquad t_{n} \sim n^{2/d_{s}}\,,
\end{equation}
which is obtained from
(\ref{gn}), (\ref{gnas}) and (\ref{sn}). Indeed, for (\ref{gn0}) to be
the limit of (\ref{gneps}) as $\varepsilon\downarrow 0$, one has to
identify $a=d_{s}/2$, and the diverging $\varepsilon$-dependencies in
$f_{3}$ have to cancel each other in order to yield a non-trivial
limit. This is only possible if $f_{3}(x,y) \sim f_{2}(y/x^{\alpha})$ as
$x,y \to 0$ with some exponent $\alpha$ which guarantees
$t^{*}(\varepsilon) \sim
\bigl(n^{*}(\varepsilon)\bigr)^{\alpha}$. Consequently, we also get
the relation $t_{n} \sim n^{\alpha}$. Hence, $\alpha = 2/d_{s}$ by
(\ref{gn0}), and we conclude $t^{*}(\varepsilon) \sim
\varepsilon^{-z}$ with $z = 2/(d_{s}\sigma)$.
Moreover, expressing $n^{*}(\varepsilon)$ in terms of
$t^{*}(\varepsilon)$, the stress-relaxation function (\ref{gint})
takes on the form 
\begin{equation}
  G(t,\varepsilon) \sim t^{-\Delta} \;
  (t/t^{*}(\varepsilon))^{\Delta -d_{s}/2} 
  \int_{0}^{\infty} \!\rmd x\; x^{1-\tau}
  f_{1}(x) \,  f_{3}(x, t/t^{*}(\varepsilon)) 
\end{equation}
for $\varepsilon \ll 1$ and $t \to\infty$ with  
$\Delta = (\tau -1)d_{s}/2$. This completes the derivation of
(\ref{deltaz}). 

\section{Discussion and outlook} \label{outlook}
%

Table~\ref{expo} summarizes the exact critical exponents for stress
relaxation in the Rouse model for gelling polymers below and at the gel
point. The exponents differ from the predictions $k=2\nu -\beta$ and
$\Delta = d\nu /(d\nu +k)$ of earlier scaling arguments \cite{DuDe87,%
  MaAd88,CoGi93,Gen78,MaAd89,RuCo89}. What is the reason for this
discrepancy? The scaling arguments  
involve the Hausdorff fractal dimension $d_{f} := d -\beta/\nu$ of
\emph{rigid} percolation clusters at $c_{\mathrm{crit}}$. Rouse clusters, 
however, are thermally stabilized, Gaussian phantom clusters with the
Hausdorff fractal dimension $d_{f}^{(\mathrm{G})} := 2d_{s} /(2-d_{s})$
\cite{Cat84,Vil88,SoBl95}, which is different from $d_{f}$ in space
dimensions below the upper critical dimension $d_{u}=6$. Indeed, if
one replaces  $d_{f}$ by 
$d_{f}^{(\mathrm{G})}$ in these scaling arguments, as one should 
consistently do in a Rouse description, the results will
coincide with Table~\ref{expo}. 

Since the scaling
relations $k=2\nu -\beta$ and $\Delta = d\nu /(d\nu +k)$ involve the
Hausdorff fractal dimension $d_{f}$ of rigid percolation clusters, it
is sometimes argued that they describe the behaviour of a more
realistic model, which, in addition to the interactions of the Rouse
model, accounts for excluded-volume effects, too, see
e.g. \cite{RuCo89}. To the best of the author's knowledge this claim
has not been verified yet by strict analytical arguments within
a microscopic model. One may even have doubts whether this claim
is generally true: Extensive molecular-dynamics simulations
\cite{VePl01} of a system of crosslinked soft spheres in three
dimensions, with a cluster ensemble as in the present Paper and an
additional strongly repulsive interaction at short distances,
yield the values  
$k\approx 0.7$ and $\Delta\approx 0.75$, which are remarkably close to
the predictions of the Rouse model for gelling polymers, see
Table~\ref{expo}. On the other hand, simulations of the
bond-fluctuation model in \cite{GaAr00} imply $k\approx 1.3$ and are thus
in favour of the claim. However, the viscosity is not measured directly in
these latter simulations. Rather it is derived from the scaling of diffusion
constants and an additional scaling assumption that may be questioned
\cite{VePl01}. Hence, it is an open problem to what extent the
critical exponents of Table~\ref{expo} are 
modified by excluded-volume interactions. 


\ack{
  My sincere thanks go to Henning L\"owe and Annette Zippelius for 
  both enlightening discussions and a critical reading of the
  manuscript. Partial financial support by SFB~602 of the
  Deutsche Forschungsgemeinschaft is acknowledged.
}

%
%


\section*{References}
\begin{thebibliography}{99}
\frenchspacing 

\bibitem{StCo82} 
  Stauffer D, Coniglio A and  Adam M 1982 \emph{Adv. Polym. Sci.} 
  \textbf{44} 103 
\bibitem{DaLa90}
  Daoud M and Lapp A 1990 \emph{J. Phys. Cond. Matter} \textbf{2} 4021
\bibitem{MaAd91} 
  Martin J E and Adolf D 1991 \emph{Annu. Rev. Phys. Chem.} \textbf{42} 311 
\bibitem{WiMo97} 
  Winter H H and Mours M 1997 \emph{Adv. Polym. Sci.} \textbf{134} 165 
\bibitem{AdDe81} 
  Adam M, Delsanti M, Durand D, Hild G and Munch J P 1981 
  \emph{Pure Appl. Chem.} \textbf{53} 1489 
\bibitem{DuDe87} 
  Durand D, Delsanti M, Adam M and Luck J M 1987 \emph{Europhys. Lett.}
  \textbf{3} 297
\bibitem{MaAd88}
  Martin J E, Adolf D and Wilcoxon J P 1988  \emph{Phys. Rev. Lett.}
  \textbf{61} 2620
\bibitem{AdMa90}
  Adolf D and Martin J E 1990 \emph{Macromolecules} \textbf{23} 3700 
\bibitem{DeBo93}
  Devreux F, Boilot J P, Chaput F, Malier L and Axelos M A V 1993
  \emph{Phys. Rev. E} \textbf{47} 2689 
\bibitem{CoGi93}
  Colby R H, Gillmor J R and Rubinstein M 1993 \emph{Phys. Rev. E}
  \textbf{48} 3712
\bibitem{VlCh98}
  Vlassopoulos D, Chira I, Loppinet B and McGrail P T 1998
  \emph{Rheol. Acta} \textbf{37} 614 
\bibitem{ToFa01}
  Tordjeman P, Fargette C and Mutin P H 2001 \emph{J. Rheol.} \textbf{45} 995 
\bibitem{Sta76} 
  Stauffer D 1976 \emph{J. Chem. Soc. Faraday Trans. II} \textbf{72} 1354
\bibitem{Gen78} 
  de Gennes P-G 1978 \emph{Comptes Rendus Acad. Sci. (Paris)}
  \textbf{286B} 131
\bibitem{Cat84} 
  Cates M E 1984 \emph{Phys. Rev. Lett.} \textbf{53} 926 
  \nonum Cates M E 1985 \emph{J. Physique (France)} \textbf{46} 1059 
\bibitem{MaAd89} 
  Martin J E, Adolf D and Wilcoxon J P 1989 \emph{Phys. Rev. A} 
  \textbf{39} 1325 
\bibitem{RuZu90} 
  Rubinstein M, Zurek S, McLeish T C B and Ball R C 1990
  \emph{J. Physique (France)} {\bf 51} 757 
\bibitem{BrGo97} 
  Broderix K, Goldbart P M and Zippelius A 1997 \emph{Phys. Rev. Lett.}
  \textbf{79} 3688 
\bibitem{ZiGr98}
  Zilman A G and Granek R 1998 \emph{Phys. Rev. E} \textbf{58} R2725
\bibitem{BrLo99}
  Broderix K, L\"owe H, M\"uller P and Zippelius A 1999 
  \emph{Europhys. Lett.} \textbf{48} 421
\bibitem{BrLo01a}
  Broderix K, L\"owe H, M\"uller P and Zippelius A 2001 
  \emph{Phys. Rev. E} \textbf{63} 01151
\bibitem{BrAs01} 
  Broderix K, Aspelmeier T, Hartmann A K and Zippelius A 2001 
  \emph{Phys. Rev. E} \textbf{64} 021404 
\bibitem{BrLo01b} 
  Broderix K, L\"owe H, M\"uller P and Zippelius A 2001 
  \emph{Physica A} \textbf{302} 279 
\bibitem{BrMu02} 
  Broderix K, M\"uller P and Zippelius A 2002 \emph{Phys. Rev. E}
  \textbf{65} 041505 
\bibitem{GaAr00}
  Del Gado E, de Arcangelis L and Coniglio A 2000 \emph{Eur. Phys. J. E}
  \textbf{2} 359 
\bibitem{VePl01} 
  Vernon D, Plischke M and Jo\'os B 2001 \emph{Phys. Rev. E} \textbf{64}
  031505 
\bibitem{Fer80}
  Ferry J D 1980 \emph{Viscoelastic Properties of Polymers} 2nd ed
  (New York: Wiley) 
\bibitem{Rou53}
  Rouse P E 1953 \emph{J. Chem. Phys.} \textbf{21} 1272
\bibitem{DoEd86} 
  Doi M and Edwards S F 1986 \emph{The Theory of Polymer Dynamics} 
  (Oxford: Clarendon Press)  
\bibitem{BiCu87} 
  Bird R B, Curtiss C F, Armstrong R C and Hassager O 1987  
  \emph{Dynamics of Polymeric Liquids} vol~2, 2nd ed 
  (New York: Wiley)
\bibitem{Vil88}
  Vilgis T A 1988 \emph{Physica A} \textbf{153} 341 
\bibitem{SoSc93} 
  Sommer J-U, Schulz M and Trautenberg H L 1993 \emph{J. Chem. Phys.}
  \textbf{98} 7515 
\bibitem{SoBl95}
  Sommer J-U and Blumen A 1995 \emph{\JPA} \textbf{28} 6669
\bibitem{BlJu02}
  Blumen A and Jurjiu A 2002 \emph{J. Chem. Phys.} \textbf{116} 2636 
\bibitem{RuCo89}
  Rubinstein M, Colby R H and Gillmor J R 1989 \emph{Space-Time
    Organization in Macromolecular Fluids} ed F Tanaka,
    M Doi and T Ohta (New York: Springer)
\bibitem{StAh94}
  Stauffer D and Aharony A 1994 \emph{Introduction to Percolation Theory}
  revised 2nd ed (London: Taylor and Francis)
\bibitem{BuHa96}
  Bunde A and Havlin S 1996 \emph{Fractals and Disordered Systems}
  ed A Bunde and S Havlin (Berlin: Springer)  pp 59, 115
\bibitem{AlOr82}
  Alexander S and Orbach R 1982 \emph{J. Physique (France)}
  \textbf{43} L-625 
\bibitem{Gen79} 
  de Gennes P-G 1979 \emph{J. Physique (France) Lett.} \textbf{40}, L-197  
\bibitem{HaZw77} 
  Haan S W and Zwanzig R 1977 \emph{\JPA} \textbf{10} 1547 
\bibitem{Gen76}
  de Gennes P-G 1976 \emph{La Recherche} \textbf{7} 919
\bibitem{HaBe02}
  Havlin S and Ben-Avraham D 2002 \emph{Adv. Phys.} \textbf{51} 187   
\end{thebibliography}
\end{document}